# Data-driven yaw misalignment correction for utility-scale wind turbines

Linyue Gao[1,3], Jiarong Hong[1, 2, a]

**AFFILIATIONS**
[1] St. Anthony Falls Laboratory, University of Minnesota, Minneapolis, MN 55414, USA
[2] Department of Mechanical Engineering, University of Minnesota, Minneapolis, MN 55455, USA
[3] California State University, Sacramento, Sacramento, CA 95819, USA.
[a] Author to whom correspondence should be addressed: jhong@umn.edu (J. Hong).

**ABSTRACT**

In recent years, wind turbine yaw misalignment that tends to degrade the turbine power production and impact the blade fatigue loads raises more attention along with the rapid development of large-scale wind turbines. The state-of-the-art correction methods require additional instruments such as LiDAR to provide the ground truths and are not suitable for long-term operation and large-scale implementation due to the high costs. In the present study, we propose a framework that enables the effective and efficient detection and correction of static and dynamic yaw errors by using only turbine SCADA data, suitable for a low-cost regular inspection for large-scale wind farms in onshore, coastal, and offshore sites. This framework includes a short-period data collection of the turbine operating under multiple static yaw errors, a data mining correction for the static yaw error, and ultra-short-term dynamic yaw error forecasts with machine learning algorithms. Three regression algorithms, i.e., linear, support vector machine, and random forest, and a hybrid model based on the average prediction of the three, have been tested for dynamic yaw error prediction and compared using the field measurement data from a 2.5 MW turbine. For the data collected in the present study, the hybrid method shows the best performance and can reduce total yaw error by up to 85% (on average of 71%) compared to the cases without static and dynamic yaw error corrections. In addition, we have tested the transferability of the proposed method in the application of detecting other static and dynamic yaw errors.

## I. INTRODUCTION

As one of the most promising renewable energy providers, wind energy has a global cumulative power capacity of up to 743 GW by the end of 2020 [1]. Currently, most utility-scale wind turbines are horizontal-axis turbines and designed to be equipped with yaw systems to keep the turbines aligned with the inflow wind for maximum energy capture [2,3]. However, in practice, there is a misalignment between the inflow wind direction and turbine nacelle orientation, named "yaw misalignment" or "yaw error". The existence of yaw error lowers the efficiency and reliability of wind turbines and thus results in high operational and maintenance (O & M) costs in the long term [4]. Specifically, yaw error significantly reduces the power capture and increases fatigue load acting on turbine blades. Besides, yaw error can also lead to inaccurate power forecasts and reduce the economy of power grid dispatching [5]. Therefore, it is essential to eliminate the negative effect of yaw error.

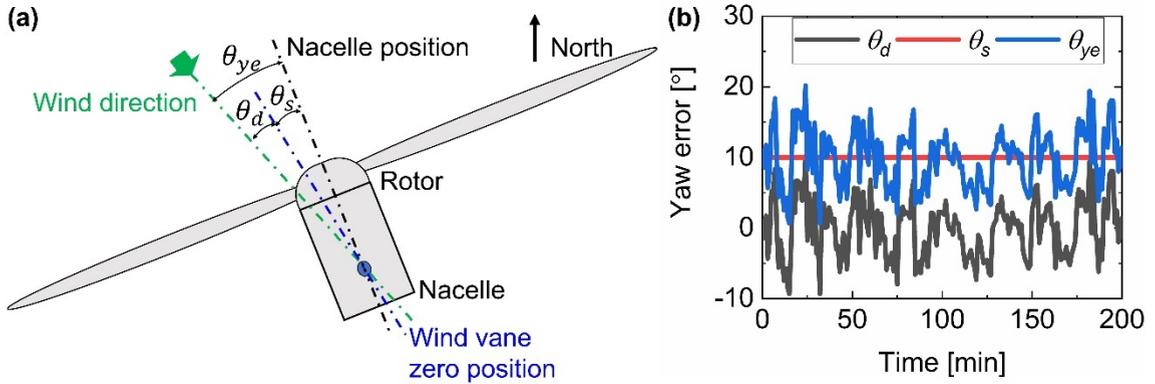

**FIG. 1**. (a) A schematic diagram illustrating an example of total yaw error $\theta_{ye}$, and its two components of static yaw error $\theta_s$, and dynamic yaw error $\theta_d$. Note that $\theta_s$ and $\theta_d$ are not always in the same direction. (b) A sample time series of $\theta_{ye}$, $\theta_s$, and $\theta_d$ collected with a field measurement of a 2.5 MW wind turbine at Eolos Wind Energy Research Station at the University of Minnesota in 2017. The temporal resolution of the data is 1 min. In this test, $\theta_s$ is set to 10° (clockwise from top) in the yaw controller, matching the mean value of $\theta_{ye}$ in the 200-min sample period. The $\theta_d$ can be seen as the fluctuating component of $\theta_{ye}$.

The yaw error of a utility-scale wind turbine generally consists of two components, i.e., static yaw error and dynamic yaw error, as shown in Fig. 1. Static yaw error (denoted as $\theta_s$), also called the zero-point shift in some investigations [6,7], is the misalignment between real nacelle position and nacelle vane zero position, i.e., measured nacelle position in turbine supervisory control and data acquisition (SCADA) system. Such angle deviation is an inherent error due to improper installation or poor calibration of the nacelle vane and its orientation shift during long-term operation [8]. The occurrence of static error is a common phenomenon [9], and the diagnosis of a substantial error of 4° by Fleming et al. [10] in 2014 starts arousing wide attention to this issue. Such error is further reported to range from a few degrees to as much as 35° [11]. Still, up to now, its examination is not included in the regular inspection because the manual examination is labor-intensive and unable to be conducted for all turbines in a wind farm. The state-of-the-art methods detect the bias in the wind direction measurements during a certain period between turbine nacelle vane and other instruments, such as forward direction wind Laser Imaging Detection and Ranging (LiDAR) on turbine nacelles [10,12], ground LiDAR [9], spinner anemometer at hub [13], and met tower anemometer right upstream the turbine. However, additional installation of such instruments is expensive, which creates a trade-off between their costs and benefits, and is not affordable for the periodical examination for hundreds or even a small number of wind turbines in a wind farm [11].

Consequently, low-cost approaches without involving any additional equipment are highly desired. In recent years, a few investigations have been performed using data-driven methods to address this issue. One category is to use the relative relations among neighboring turbines to identify the anomalous turbine and assume the selected turbine owns substantial static yaw error [6,14,15]. Specifically, Astolfi et al. [14] detected one of the six nearby wind turbines sited in a flat wind farm possessing static yaw error based on its anomalous statistics in relative nacelle direction. The same group also used the power coefficient curve [6] and the rotor speed curve [15] as indicators. However, such approaches focus on detecting the presence of static yaw error, and they are only applicable to wind farms in flat terrains where local winds are not strongly affected by terrain-induced flow disturbance. The other category is to use the relative relation between the nominal turbine performance with zero static yaw and measured turbine performance in practice, i.e., performance degradation. In studies [6] and [15], Astolfi et al. found shifts of the highest values in power coefficient and rotor speed as a function of the yaw angle from zero to 8° and -5°, respectively, and such shifts implied the corresponding static yaw errors. A similar procedure has been applied to the indicator of power output versus yaw error curve by Pei et al.[7]. Compared to the ground truths of static yaw error obtained with manual inspections by sending technicians to the site to inspect the wind vanes mounted the top on turbine nacelles (with a statistical accuracy of up to 0.1°) [16], the above quantitative estimations exhibit high uncertainties (i.e., >20%), mainly associated with the embedded dynamic yaw errors in the datasets and the statistical interval size of yaw error (1° or 2°). Instead, the utilization of the theoretical cosine-cubic relation between power and yaw error (i.e., $P \propto cos^3(\theta_{ye})$) reduces the uncertainty levels by eliminating such influences of dynamic yaw and interval size. Bao et al. [8] fitted the cosine-cubic relation using a nonlinear least square method for a dataset simulated with GH Bladed (i.e., an integrated software developed by Garrad Hassan for wind turbine performance and loading simulations) and reported a 4% prediction error in static yaw. Such simulation simplifies the real situations to some extent and requires field measurements for further validation. Multiple fitting methods tested by Xue et al. [17] on turbine SCADA data show very close prediction results, suggesting the type of fitting method has limited influence on the outcome. However, this study lacks the ground truth of the static error for the evaluation of prediction uncertainties. More importantly, it should be noted that the cosine-cubic relation of power and yaw error could, in fact, be cosine-squared, or even cosine relying on the aeroelastic properties and control systems of the turbine [10] and site-dependent inflow wind fields [18,19], which suggests that such a relationship should be evaluated using field measurement data before the implementation to the yaw error detection. In brief, the limitations in the previous studies on the static yaw detection include the requirement of additional devices (LiDAR, spinner anemometer, met tower, and so forth), low accuracy in the quantitative estimation, and lack of field measurement validation.

The other component of yaw error is dynamic yaw error (denoted as $\theta_d$), defined as the instantaneous misalignment between wind direction and zero position of the wind vane mounted on the back of turbine nacelle (i.e., the measured nacelle position in SCADA), as shown in Fig. 1. Such angle deviation is mainly caused by the delayed response of yaw control or the uncertainties in wind direction measurement at nacelle associated with flow distortion [10]. Most studies focus on utilizing wind direction forecasts to mitigate the negative influence caused by such delayed response [9,20–22]. For example, Choi et al. [9] applied multiple machine learning-based (ML-based) models (including linear regression, gradient boosting, and random forest) to predict wind direction and presented 4%-13% error reduction rates after the correction of the dynamic component. However, it is challenging for the general algorithms to predict the 360°-span of wind direction accurately. These algorithms require additional transformation to convert the angular form of wind direction into a linear form [23], introducing other uncertainties in the forecast. In fact, based on turbine yaw control systems, most utility-scale turbines can align with the inflow according to wind direction measurements by their nacelle wind vanes/anemometers with instantaneous $\theta_d$ changing mostly between -10° to 10° [16]. The direct forecast of dynamic yaw error is a promising option to mitigate the influence associated with the large-span issue in the wind direction forecast.

To address the above limitations, we propose a comprehensive framework that can effectively and efficiently detect and correct yaw error using the turbine SCADA data without involving additional instruments. Section II and Section III present the proposed data-driven approaches to detect the static and dynamic yaw errors, respectively. Section IV illustrates the overall framework of how to apply the data-driven approaches to a practical application in a wind farm without using additional instruments. In Section V, such framework is evaluated by the field measurements at Eolos Wind Energy Research Station of the University of Minnesota (referred to as Eolos station hereafter), followed by a brief discussion on transferability of the proposed framework. Section VI summarizes the research findings and the potential implementation of the proposed framework for turbine inspection in wind farms.

## II. STATIC YAW ERROR DETECTION

### A. Power-yaw relationship with a benchmarking turbine test

The existence of yaw error tends to reduce the inflow wind speed relative to the rotor swept area due to the flow distortion. Such reduction contributes to the degradation of the turbine power output (denoted as P(t)) following a cosine law, as described in the model of turbine power in yaw misalignment in Eq. (1).

$$P(t) = \frac{1}{2}\rho(t)AC_p\{V^3(t)cos[\theta_{ye}(t)]^\alpha\} \tag{1}$$

where $\rho(t)$ is the air density, $A$ is turbine rotor sweep area, $C_p$ is power coefficient varying with the turbine operational regions, $V(t)$ is the inflow wind speed at turbine hub height, and $\alpha$ is the exponent of the cosine relationship between yaw error and turbine power.

Static yaw error ($\theta_s$) is one of the components of total yaw error ($\theta_{ye}(t)$), as given in Eq. (2), responsible for the power reduction. Our proposed method for static yaw error detection roots in this mechanism. Since static yaw error is relatively stable in contrast to dynamic yaw error, we assume that static yaw error is time-independent within a duration significantly longer than the time scale associated with the change of dynamic yaw error.

$$\theta_{ye}(t) = \theta_s + \theta_d(t) \qquad (2)$$

where $\theta_d(t)$ refers to time-varying dynamic yaw.

To simply Eq. (1), we use $P_0(t)$ to represent reference power output with zero static yaw misalignment, as shown in Eq. (3).

$$P_0(t) = \frac{1}{2}\rho(t)AC_p V^3(t) \qquad (3)$$

In Eq. (1), theoretically, the exponent $\alpha$ equals 3, the same as the exponent for $V$, which is widely used in previous studies [6,7]. However, in practical applications, this exponent on $cos\theta_{ye}$ is sensitive to the aeroelastic properties of the turbine blades [10] and subject to the turbine control systems [24].

To leverage the impact of yaw error on turbine power for static error detection, the first step is to find this power-yaw relationship. A benchmarking turbine (randomly selected or the one with less fault reporting in a wind farm) with recent manual inspection is selected for the collection of the datasets with intended static yaw errors (i.e., $\theta_s$, ground truth) by setting corresponding offsets in the turbine yaw controller. The exponent $\alpha$ can be extracted from the measured data from the benchmarking turbine test using a nonlinear least-square fitting method subject to Eq. (4).

$$\min \sum[P_0(t) - P(t)/\cos(\theta_d(t) + \theta_s)^\alpha]^2 \qquad (4)$$

In the calculation using the measured dataset, $\theta_d(t)$ is the angle deviation between the measured wind direction $\theta_{WD}(t)$ and measured nacelle direction $\theta_{ND}(t)$, i.e., $\theta_d(t) = \theta_{WD}(t) - \theta_{ND}(t)$, available in the turbine SCADA system. $\theta_s$ is a constant, referring to the intended static yaw error, i.e., the bias we set in the turbine yaw controller for the benchmarking test. $P(t)$ is the measured active power recorded in the SCADA system.

The reference power $P_0(t)$ is difficult to be derived from Eq. (3) due to the varying value of $C_p$. Instead, it can be calculated via the turbine power curve (Fig. 2) and the measured wind speed data prepared in the form of time series according to Eq. (5). In Region II [25], the turbine operates in a variable-speed scheme to capture as much power as possible from the wind with a small constant or even zero pitch angle, while the turbine falls in Region III with a variable-pitch scheme to remain safe operations with constant rated power output. Noted that to eliminate the impact of blade pitching on the turbine power output, only the data corresponding to the turbine operating in the variable-speed scheme, i.e., Region II, are kept in the estimation of the exponent on $cos\theta_{ye}$ in this study. In addition, the ninth-order polynomial fitting performs best among different-degree fittings in this region [26] and is used in the present study. The turbine power curve is designed with a standard air density of $\rho^*$, and based on the proportional relation between power and air density given in Eq. (3), we can eliminate the air density effect on the $P_0(t)$ estimation by introducing a factor of $\rho(t)/\rho^*$.

$$P_0(t) = (c_l V(t)^l + c_{l-1} V(t)^{l-1} + \cdots + c_1 V(t) + c_0) \cdot \rho(t)/\rho^* \qquad Region\ II \qquad (5)$$

where $c_1, \dots, c_l$ are regression constants and $l$ is the degree of the model, $P_{rated}$ is the turbine rated power, $\rho^*$ is the standard air density used for the designed power curve, i.e., $\rho^* = 1.225\ kg/m^3$, and $\rho(t)$ is the real-time recorded air density in the turbine SCADA system.

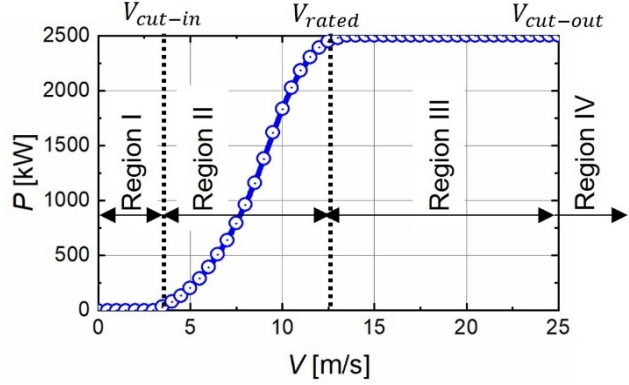

**FIG. 2**. Wind turbine power curve and operational regions, including Region I ($V < V_{cut-in}$, cut-in wind speed), Region II ($V_{cut-in} \leq V \leq V_{rated}$, rated wind speed), Region III ($V_{rated} \leq V \leq V_{cut-out}$), and Region IV ($V > V_{cut-out}$, cut-out wind speed).

### B. Static yaw error detection

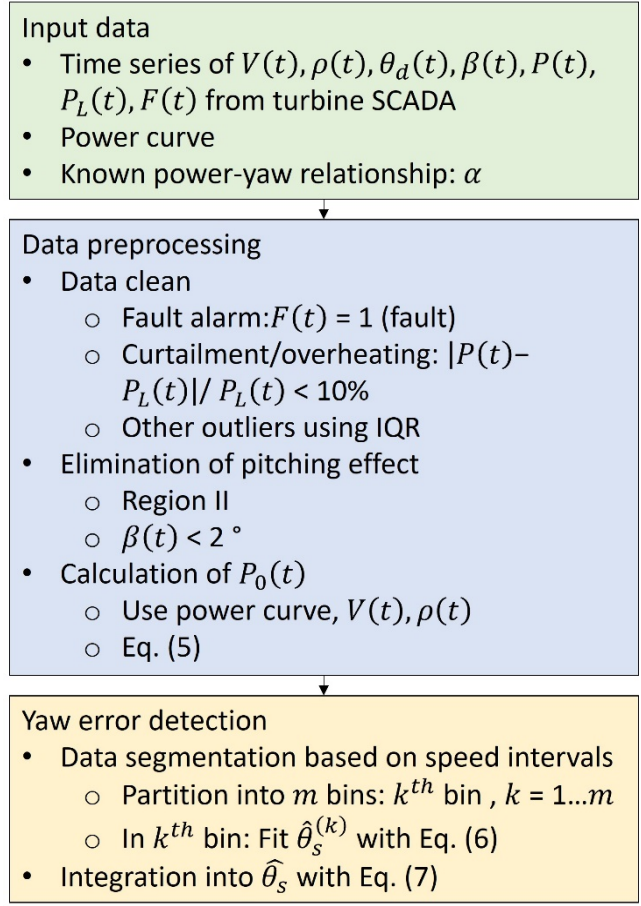

**FIG. 3**. Schematic of turbine static yaw error detection.

The known power-yaw relationship ($\alpha$) derived from the benchmarking turbine test is employed to detect the static yaw errors for the other turbines in the same wind farm. Fig. 3 shows the schematic of the turbine static yaw error detection process. The input information consists of the measured datasets of hub-height wind speed $V(t)$, air density $\rho(t)$, pitch angle $\beta(t)$, dynamic yaw error $\theta_d(t)$, power $P(t)$, power limit $P_L(t)$, fault code $F(t)$, and the turbine power curve. The data labeled with fault alarms (fault: $F(t) = 1$ and no-fault: $F(t) = 0$) and affected by the grid curtailment and overheating issues ($|P(t)\text{-}P_L(t)|/ P_L(t) < 10\%$) are discarded. The remaining outliers in the data are further removed using the interquartile range (IQR) method [27]. To eliminate the blade pitching effect on power deviation, only the data in Region II with a pitch angle lower than 2° are kept as the preprocessed dataset for next-step analysis. The reference power with zero yaw error $P_0(t)$ needs to be calculated with the sequences of $V(t)$, $\rho(t)$ and turbine power curve.

The preprocessed dataset is partitioned into $m$ bins based on the wind speed intervals to eliminate the influence of wind speed on turbine power production. Such segmentation approach has been widely used by many researchers for similar purposes [7,12]. Based on the bin size, the actual number of bins depends on the range of wind speed for the dataset (within Region II). In each wind speed bin, the nonlinear least-square method is adopted to fit static yaw error, subject to Eq. (6). Note that the power-yaw relationship ($\alpha$) in Eq. (6) is a known quantity derived from the benchmarking test using the methods mentioned in Section IIA.

$$\min \sum \{P_0^{(k)}(t) - P(t)/\cos [\hat{\theta}_s^{(k)} + \theta_d^{(k)}(t)]^\alpha\}^2 \tag{6}$$

where $k = 1, \ldots, m$ and superscript represents the $k^{\text{th}}$ bin. $\hat{\theta}_s^{(k)}$ represents the estimated static yaw error in the $k^{th}$ bin.

The estimated results in different bins are then integrated into the final estimated static yaw error for the correction by using their arithmetic mean, as expressed in Eq. (7).

$$\hat{\theta}_s = \frac{1}{m}\sum_{k=1}^{m} \hat{\theta}_s^{(k)} \tag{7}$$

The performance of the proposed approach is evaluated using the metric of relative mean absolute error (RMAE), as defined in Eq. (8).

$$RMAE = \frac{\hat{\theta}_s - \theta_s}{\theta_s} \times 100\% \tag{8}$$

## III. DYNAMIC YAW ERROR PREDICTION

### A. ML-based algorithms

We predict time-varying values of dynamic yaw error from the historical turbine SCADA data using ML-based algorithms. Such ML-based tools perform well when the relationship between the inputs and output is not clear and have been successfully utilized to forecast turbine features, such as power output [28–33]. In the present study, we use three popular ML-based algorithms for dynamic yaw error prediction, namely linear regression, support vector machine (SVM) regression, and random forest (RF) regression. To optimize the hyperparameters for each algorithm, we use the trial-and-error approach [33] with 10% of the data to run algorithms with different settings and select the best performer, as shown in Appendix A of the supplementary material. Detailed information on the three algorithms is summarized below. In addition, we also introduce a hybrid model based on the above three ML-based algorithms by averaging the prediction results from them.

First, linear regression algorithms are suitable for high-dimensional, full, or sparse predictor data, as used in the present study, for fast prediction, as shown in Eq. (9). Regularized SVM and least-squares regression are two widely used linear regression algorithms, whose loss functions are shown in Eq. (10) and Eq. (11), respectively. With the optimization of the hyperparameters (Fig. S1 in the supplementary material), least-squares regression exhibits a better performance than the regularized SVM and thus is used here to train the linear regression model. The objective function minimization technique is Broyden-Fletcher-Goldfarb-Shanno quasi-Newton algorithm (BFGS) [34] with L2 (ridge) regularization.

$$f(x) = x\beta + b \tag{9}$$

where $x$ is an observation from predictor variables, $\beta$ is a vector of predictor coefficients, and $b$ is the scalar bias.

$$\iota[y, f(x)] = \max[0, |y - f(x)| - \varepsilon] \tag{10}$$

where $y$ is the response data, i.e., ground truth for the prediction results, and $\varepsilon$ represents half the width of epsilon-insensitive band, i.e., precision.

$$\iota[y, f(x)] = \frac{1}{2}[y - f(x)]^{1/2} \tag{11}$$

Second, SVM [35] is a widely-used algorithm for regression problems for time series of wind data, enabling fast and robust prediction. SVM algorithm for regression uses the training data to fit an appropriate hyperplane within a specific margin. The primal formula of SVM is presented in Eq. (12). In this study, the Gaussian kernel function (see Eq. (13)) is used with optimization of sequential minimal optimization (SMO) to achieve the nonlinear regression.

$$\frac{1}{2}\|\beta\|^2 + C\sum_{n=1}^{N}(\xi_n + \xi_n^*) \text{ subject to } \begin{cases} y_n - (\langle x_n, \beta\rangle + b) \leq \varepsilon + \xi_n \\ \langle x_n, \beta\rangle + b - y_n \leq \varepsilon + \xi_n \\ \xi_n, \xi_n^* \geq 0 \end{cases} \tag{12}$$

where $\xi_n$ and $\xi_n^*$ are slack variables for each point, i.e., "soft margin" for SVM. Box constraint $C$ is a positive numerical value that controls the penalty imposed on observations outside the epsilon margin ($\varepsilon=1.69$ in the present study) and helps prevent overfitting ($C = 0.0010019$ in the present study).

$$G(x_j, x_k) = \exp(-\|x_j - x_k\|^2) \tag{13}$$

Third, ensemble ML-based methods take advantage of their benefits in improving the average prediction performance over any contributing member in the ensemble. With the optimization of the hyperparameters (Fig. S3 in the supplementary material), one of the popular ensemble methods of decision trees, i.e., RF, is used in the present study. RF [36] is a popular ensembled method of decision trees with a basic principle of bagging, i.e., the bootstrap aggregation that randomly selects a portion of data from the training set to fit a tree and then averages the outputs of all trees [37]. In this study, the number of learning cycles is selected as 400, and the minimum number of leads required to split an internal node is 4.

## B. Dynamic yaw error forecast using ML-based algorithms

The forecast of dynamic yaw error is used to mitigate the misalignment effect for real-time control optimization. It requires an ultra-short-term forecast (1-min ahead) along with a high time resolution (1 min in this study). The forecast dynamic yaw error (denoted as $\hat{\theta}_d(t+1)$, and "1" corresponds to 1 min in this study) is modeled based on three features selected from the high dimensional SCADA data using the grey relational analysis (GRA) [38] (see Appendix B of the supplementary material), including turbine power ($P$), wind speed ($V$), wind direction ($\theta_{WD}$) as expressed in Eq. (14).

$$\hat{\theta}_d(t+1) = f(P(t), \theta_{WD}(t), V(t)) \tag{14}$$

The dynamic yaw error significantly depends on its past values within 5 min, and thus the input features are extended to Eq. (15)

$$\hat{\theta}_d(t+1) = f(\theta_d(t), \theta_d(t-1)\ldots\theta_d(t-4), P(t), \theta_{WD}(t), V(t)) \tag{15}$$

where $\theta_d(t), \ldots, \theta_d(t-4)$ represent the current dynamic yaw errors and four past observations within 5 min.

The entire observations are split into training and testing sets with a ratio of 4:1, which is commonly used for wind energy forecasts [39]. The three candidate ML-based algorithms are trained using the training sets. The performance of the algorithms is evaluated based on their forecasted data sequences ($i = 1, \ldots, n$) using two metrics, i.e., mean absolute error (MAE) and root mean square error (RMSE), as given in Eq. (16) and (17). These two metrics are widely used for the performance evaluation for the ML-based algorithms for short-term forecasts [40,41].

$$MAE = \sum_{i=1}^{n} |\theta_{d,i} - \hat{\theta}_{d,i}|/n \tag{16}$$

$$RMSE = \sqrt{\sum_{i=1}^{n} |\theta_{d,i} - \hat{\theta}_{d,i}|^2 / n} \tag{17}$$

**IV. FRAMEWORK OF TOTAL YAW ERROR CORRECTION**

This section introduces the overall framework for the practical application of the proposed approaches in Sections II and III to utility-scale wind turbines. Fig. 4 illustrates four major parts of the framework, including data collection, data preprocessing, modeling, and correction & evaluation. First, in a wind farm, a benchmarking turbine (randomly selected or the one with less fault reporting) with recent manual inspection is selected for the collection of the datasets with intended static yaw errors (i.e., $\theta_s$, ground truth) by setting corresponding offsets in the turbine yaw controller. At least four offsets ranging from -10° (counterclockwise yaw, bird's view) to 10° (clockwise yaw) are suggested to provide adequate datasets for the modeling. The larger offsets with absolute values larger than 20° are not recommended to reduce their impact on turbine loads for the sake of turbine safety [42,43]. Additionally, a normal operation dataset (zero static yaw error confirmed with manual inspection) is used to derive a baseline power curve for comparison using the binned average power values with a size interval of wind speed of 0.5 m/s. The ideal power curve provided by the manufacturer can also be used as the baseline, which may exhibit slightly higher power outputs than the measured one as the ideal curve does not consider the turbulence effect in the field [26]. Second, the collected data from the benchmarking turbine need to be preprocessed using the data-clean criteria mentioned in Fig. 3. Such datasets are then used to determine the power-yaw relationship for the static yaw error detection approach and feature selection for the development of dynamic yaw error prediction algorithms. Third, with the derived power-yaw relationship and selected features, we can conduct the static yaw error estimation using the approach mentioned in Section II and train the ML-based models listed in Section III for the dynamic yaw error prediction. Once the models are built, only the procedures highlighted in blue in Fig. 4 need to be implemented for the detection and correction of yaw errors for other turbines. Specifically, there is no need for additional data collection or manual inspection to ensure a zero static yaw beforehand. The cleaned operation data can be directly used for the static yaw error estimation based on the predetermined power-yaw relationship and for the time-dependent dynamic yaw error forecasts using the trained models. The detected total yaw error $\hat{\theta}_{ye}$ (i.e., $\hat{\theta}_{ye}(t) = \hat{\theta}_s + \hat{\theta}_d(t)$ ) can be integrated into the turbine yaw controller as offsets for the correction of yaw misalignments. Last, the performance of the proposed methods is evaluated by comparing the ground truth with the predictions. Their statistical differences for the data sequences ($i = 1, ..., n$) represent the degree of correction, defined as correction factor of yaw error (denoted as $CF_{ye}$) in Eq. (13).

$$CF_{ye} = \left[1 - \sqrt{\sum_{i=1}^n (\hat{\theta}_{ye,i} - \theta_{ye,i})^2} \Big/ \sqrt{\sum_{i=1}^n (\theta_{ye,i})^2}\right] \times 100\% \qquad (13)$$

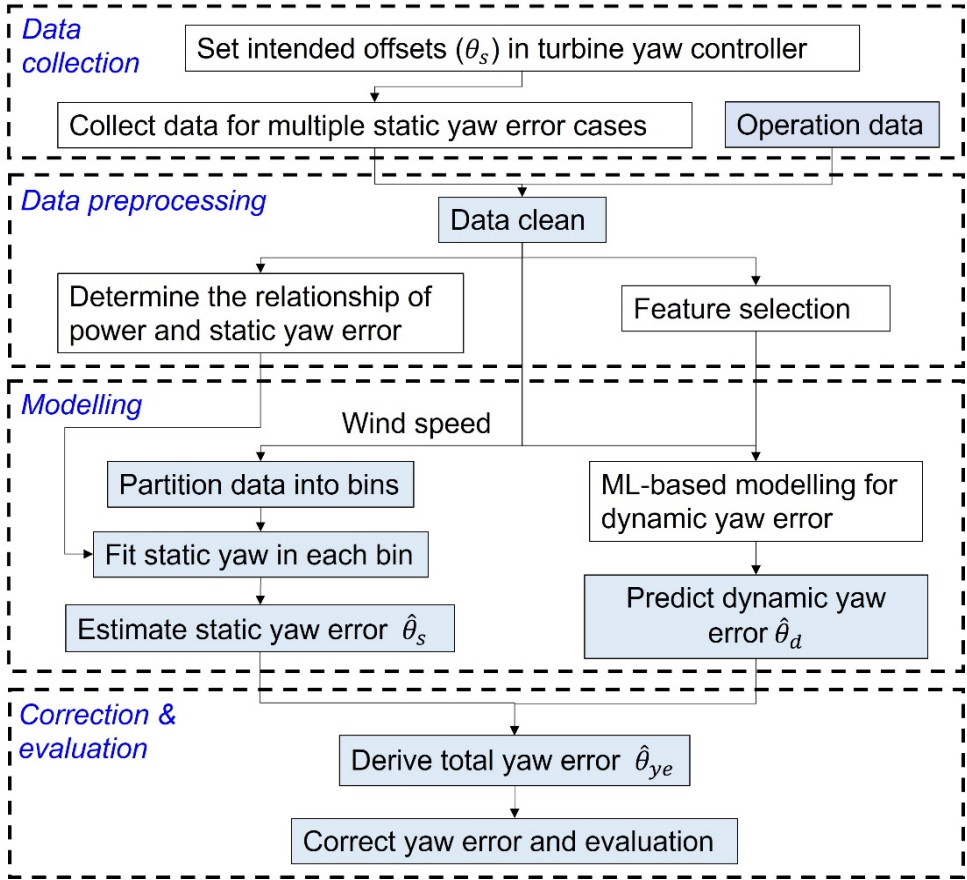

**FIG. 4**. The proposed framework for the yaw error detection and correction for a utility-scale wind turbine. Once the model is built for the benchmarking turbine, only procedures highlighted in blue are required for the implementations to other turbines in a wind farm.

## V. METHOD VALIDATION

### A. Data description

We evaluate the proposed framework with the field measurements conducted at Eolos Wind Energy Research Station, MN, USA, which hosts a 2.5 MW three-bladed horizontal-axis wind turbine (Clipper, C96). The turbine is variable-speed and variable-pitch regulated with a maximum power coefficient in operation in broad Region II (i.e., inflow wind speed $V \in [4,11]$ m/s), as shown in Fig. 2. The turbine rotor diameter and hub height are 96 m and 80 m, respectively. More detailed information on the Eolos turbine is available in our previous studies [44–46]. We collect the data in six static yaw error conditions by adding offsets to the parameter "YawErrAdj" in the yaw controller, as listed in Table I. For example, in Case#01, a 10° offset is introduced to the yaw controller to represent the condition where the turbine has a static yaw error of $\theta_s = 10°$. In the initial status (i.e., regular operation with zero static yaw error), the parameter default value is 0. Note that positive and negative signs in the intended yaw angles represent the clockwise and counterclockwise directions in the bird's view, respectively. Such intended yaw angles have been validated to be able to present the real static yaw misalignments between the turbine nacelle and the inflow wind with the comparison of the hub-height wind direction measurements by the turbine nacelle anemometer and meteorological anemometer 170 m south to the turbine at the Eolos station (see Appendix C in the supplementary material).

TABLE I. Case list of the data sets with different intended static yaw angles.

| Case# | Intended static yaw $\theta_s$ [°] | Data group number (Cleaned, 1-min averages) | Purpose |
|---|---|---|---|
| 01 | 5 | 4500 | Modeling |
| 02 | 10 | 4500 | Modeling |
| 03 | -8 | 4500 | Modeling |
| 04 | -10 | 4500 | Modeling |
| 05 | 8 | 4500 | Transferability |
| 06 | -6 | 4500 | Transferability |

The data is recorded in the turbine SCADA system with a frequency of 1 Hz. We use their 1-min averages to eliminate the high-frequency fluctuations that are significantly beyond the turbine yaw controller's capability with a slow yaw rate (about 0.5°/s for most turbines [47] for the sake of turbine operational safety). Then the data with 1-min resolutions are cleaned using the criteria for "Data clean" mentioned in Fig. 3. For each case, we tailor the dataset to 4500 data groups with 30 turbine operation features (i.e., 4500×30 data points). The exact number of data groups in each case allows a fair comparison of the performance among all six cases. The features include power, power limit, wind speed, hub speed, generator speed, rotor position, pitch angles for three blades, pitch torques for three blades, wind direction, nacelle direction, fault code, tower accelerate, air density, barometric pressure, nacelle temperature, ambient temperature, gearbox oil temperature, four high-speed gearbox bearing temperatures, hydraulic oil temperature, and four generator winding temperatures, as listed in Fig. S4 in Appendix B of the supplementary material. After the feature selection, only six remaining features are needed, power, power limit, wind speed, wind direction, nacelle direction, air density. Note that the first four cases labeled "Modeling" in Table I are used for the development of static and dynamic yaw error models, i.e., the datasets required from the benchmarking turbine. The remaining two datasets labeled "Transferability" are utilized for the transferability tests, i.e., simulating the datasets from the other two turbines.

Fig. 5 shows sample cleaned time sequences of recorded dynamic yaw error in the turbine SCADA system. The 1-min interval data retains the volatility of dynamic yaw errors associated with complex inflow structures. In the meantime, such interval allows the timely responses from the turbine yaw controller for the active correction, which is also used in [9]. As to the static yaw error, the 1-min data are smoothed to 1-h moving averages to filter out the irrelevant short-term fluctuations and ensure adequate datasets for further analysis. The standard deviation of the 1-h moving averages shown in Fig. 5 is reduced by 67% in comparison with the 1-min data. In addition, it should be noticed that the dynamic yaw errors fluctuate around zero with a small mean value within a range from -1.0° to 1.0° for all six cases. Such a trend suggests that the dynamic yaw errors exhibit no appreciable correlation with the static yaw errors, indicating that such two types of errors can be detected separately.

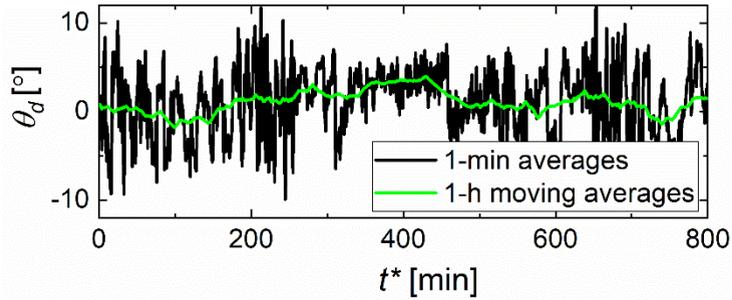

**FIG. 5.** A sample dataset of dynamic yaw errors in case#2 ($\theta_s=10°$). The equivalent time (denoted as $t^*$) corresponds to 800 valid data points for $t^*=800$ min. The green curve is obtained by smoothing the 1-min averages data to 1-h averages to filter out the irrelevant short-term fluctuations for static yaw error detection.

**B. Static yaw error detection**

Following the framework for the determination of the relationship between turbine power and yaw error, we derive the exponent of the cosine law in Eq. (1) by fitting the data of cases#01-04 with a nonlinear least-square fitting method subject to Eq. (4) with known ground truths of static errors. The exponent derived from the four cases has an average value of 3.1 with a 95% confidence in the range of 3.05 to 3.14. Fig. 6 further compares the corrected data based on different cosine laws. The one using $\alpha = 3$ matches the best with the power curve under no yaw error conditions with an averaged 1% deviation. Therefore, we use $\alpha = 3$ in the following data mining process to estimate the static yaw error. In addition, the broad Region II ($V \in [4,11]$ m/s) of the Eolos turbine consists of Region 1.5 ($V \in [4, 6.9)$ m/s), central Region II ($V \in [6.9, 9.2]$ m/s), and Region 2.5 ($V \in (9.2, 11]$). Region 1.5 and Region 2.5 are transition regions between Region I and central Region II and central Region II and Region III, respectively. The corrected data based on power-yaw relationship $\alpha = 3$ yields a good agreement with the power curve in central Region II. The correction based on $\alpha = 3$ tends to slightly (<10%) underestimate and overestimate turbine power in Region 2.5 and Region 1.5, respectively. Such errors are suggested to be associated with the lack of capability of the cosine law to capture the features due to more complex control setting changes in those transition regions.

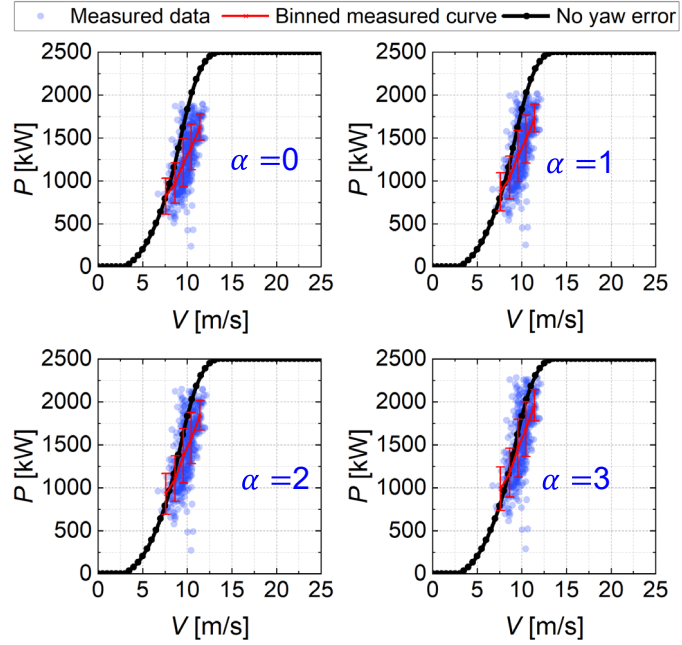

**FIG. 6**. Turbine power curve for the verification of the relation between power and yaw error. Here, the measured data are corrected using the different cosine exponents from 0 to 3. The binned curve is derived from the measured data with an interval of 1.0 m/s.

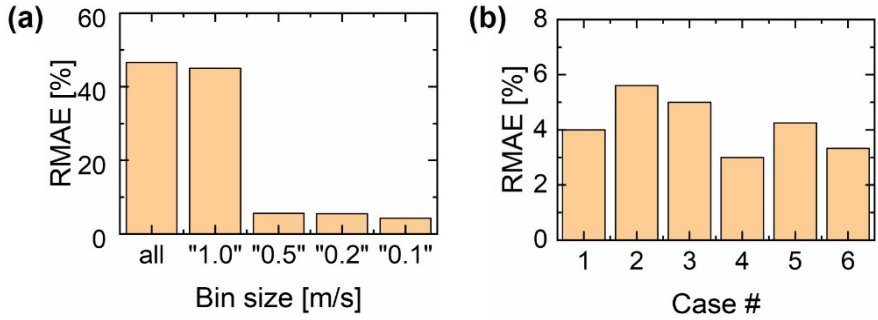

**FIG. 7**. (a) Influence of bin size on the estimation of static yaw error, and (b) the relative mean absolute error (RMAE) for all cases with a bin interval of 0.5 m/s. The term "all" in (a) presents that all the data are compiled into one bin for the estimation.

The six cleaned datasets are partitioned into bins and fitted using the nonlinear least-square method for the static yaw errors. Fig. 7(a) shows the performance of the estimation as a function of bin size using RMAE as the metric. The RMAE dramatically decreases as the bin size of wind speed reduces from 1.0 m/s to 0.5 m/s and exhibits no appreciable difference among 0.5 m/s, 0.2 m/s, and 0.1 m/s. According to such a trend, we select 0.5 m/s as the bin size for the data partition of all the cases. The RMAEs in the estimations of static yaw error are small ($< 6\%$ in general), as shown in Fig. 7(b), indicating that the proposed method has a good capability in capturing the static yaw error. It should be noted that cases#5-6 that are not included in the identification of the power-yaw relationship also show limited errors (RMAE < 5%, see Fig. 7(b)), suggesting the proposed method has good transferability.

## C. Dynamic yaw error prediction

Only the data in cases#01-04 labeled "Modeling" are used as the initial datasets for training the three regression models. The other two cases labeled "Transferability" are kept for the following turbine-turbine transferability analysis. Following the proposed framework, the first 80% of data in cases#01-04 are used as the training datasets ($n$ = 3600 per case). The rest 20% of data in cases#01-04 are set as test datasets ($n$ = 900 per case), while all data available in cases#05-06 are grouped into test datasets ($n$ = 4500 per case).

Fig. 8 shows the dynamic yaw errors forecasted using three ML-based algorithms as well as the real observations. In addition, the forecasts derived from the three trained models are further averaged to obtain a hybrid prediction. In addition, a persistence model (PM), i.e., $\hat{\theta}_d(t+1) = \theta_d(t)$, is also included here to help assess whether ML-based forecasts improve prediction accuracy. Figs. 8(a)-(d) shows the prediction results for the "Modelling" cases. In general, the forecasts derived from linear, SVM, RF, and the hybrid model can follow the general trends of the real measurements very well, particularly for the periods with benign yaw errors. For the situations in which the dynamic yaw error changes rapidly, linear, SVM, RF, and the hybrid model based on them tend to smooth such high fluctuations. Such underestimation of the changes at two adjacent time points using SVM or other ML-based models, particularly at the curves' tunning points, is also observed in the time series forecasts in other applications [48,49].

**TABLE II.** Accuracy of dynamic yaw error prediction.

| Case# | MAE [°] | | | | | | RMSE [°] | | | | | |
|---|---|---|---|---|---|---|---|---|---|---|---|---|
| | Linear | SVM | RF | Hybrid | PM | PM10 | Linear | SVM | RF | Hybrid | PM | PM10 |
| 01 | 1.9 | 2.0 | 1.9 | 1.4 | 2.2 | 2.3 | 2.4 | 2.6 | 2.4 | 1.8 | 3.0 | 3.0 |
| 02 | 2.5 | 2.5 | 2.3 | 1.8 | 2.6 | 3.6 | 3.3 | 3.4 | 3.1 | 2.4 | 3.8 | 4.5 |
| 03 | 2.7 | 2.7 | 2.3 | 1.9 | 2.9 | 3.5 | 4.1 | 4.2 | 3.6 | 2.9 | 5.0 | 4.6 |
| 04 | 3.3 | 3.4 | 3.0 | 2.3 | 3.7 | 4.2 | 4.6 | 4.5 | 4.2 | 3.2 | 5.6 | 5.2 |
| 05 | 2.7 | 2.8 | 2.4 | 1.9 | 2.9 | 3.4 | 4.0 | 3.9 | 3.6 | 2.8 | 4.8 | 4.4 |
| 06 | 3.4 | 3.4 | 3.0 | 2.4 | 3.8 | 3.9 | 5.1 | 5.0 | 4.6 | 3.6 | 6.3 | 5.3 |
| Avg | 2.7 | 2.8 | 2.5 | 2.0 | 3.0 | 3.5 | 3.9 | 3.9 | 3.6 | 2.8 | 4.7 | 4.5 |

Table II gives a more direct view of the model performance with metrics of MAE and RSME. The linear and SVM yield similar prediction errors, while RF shows a slightly better accuracy. In addition to the instantaneous persistence model (PM), a persistence model which uses a 10-min moving average of the dynamic yaw error, i.e., $\hat{\theta}_d(t+1) = \frac{1}{10}\sum_{i=1}^{10} \theta_d(t-i)$, denoted as PM10, is also introduced here as a reference. In general, PM yields lower errors in terms of MAE for most cases compared to PM10, while PM10 has a better performance with respect to RSME. In comparison to PM and PM10, the ML-based algorithms can reduce the prediction errors of MAE and RMSE by 7%~40% and 15%~23%, respectively. Although the persistence model (PM) can ultimately retain the high fluctuations in the dynamic yaw error (Fig. 8), the fixed delay (i.e., 1 min delay for PM and 1-10 min delay for PM10) is responsible for the lower accuracy compared to the forecasts derived from ML-based models. More importantly, compared to the individual ML-based algorithms, the hybrid model can significantly reduce the prediction error by 20%~29% and 22%~28% in terms of MAE and RMSE, respectively. The hybrid model takes advantage of the different performances from the independent algorithms and neutralizes their errors in different directions to achieve the best prediction.

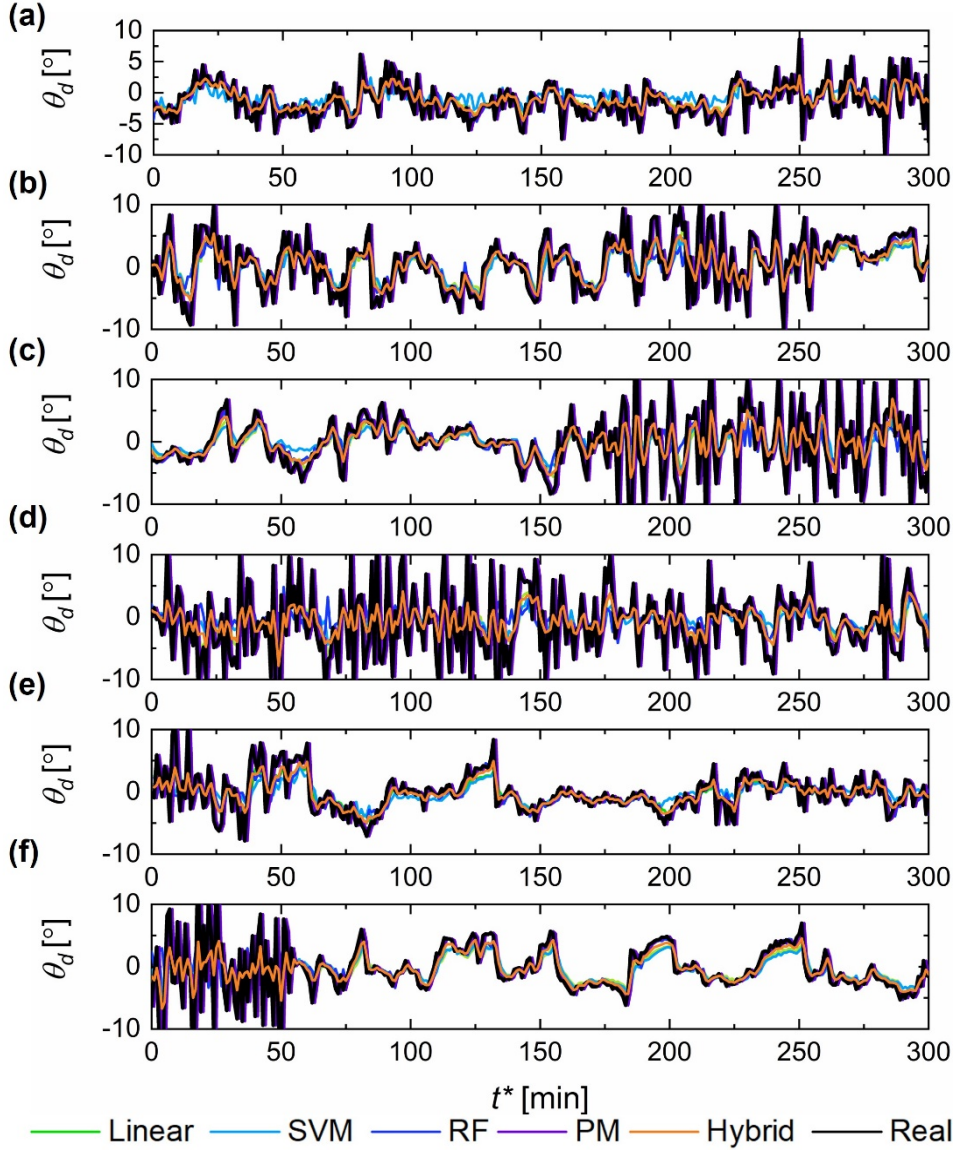

**FIG. 8**. Sample time sequences of estimated and measured dynamic yaw errors in all six cases. (a) – (f) correspond to cases#01-06, respectively. The data labeled "Real" refer to the measured dynamic yaw errors ($\theta_d(t)$) recorded in the turbine SCADA system, i.e., differences between the nacelle direction ($\theta_{ND}(t)$) and measured wind direction ($\theta_{WD}(t)$).

Figs. 8(e)-(f) present the forecasts for the two cases to test the transferability of the proposed models. Similar trends are observed in these two cases that the results derived

from linear, SVM, RF, and hybrid models have good agreements with the measured data. The forecast errors for these two cases are comparable to cases#01-04 regarding MAE and RMSE, indicating that the derived models from the benchmarking turbine have good transferability to other turbines.

In addition to the prediction accuracy of different algorithms, their computational efficiency also needs to be considered. The implementations are performed on a ThinkPad laptop with an Intel Core i7-6600U 2.60 GHz processor and 20.0 GB memory. The comparison of the testing time for different algorithms for the 200 data points of dynamic yaw errors in each case is listed in Table III. All models consume an extremely short testing time, and the maximum testing time for the three models is used as the computational time for the hybrid model. Such short testing time demonstrates the feasibility of the proposed method in predicting the dynamic yaw error and integrating such information into the turbine yaw controller for eliminating the impact of yaw error on turbine power production.

**TABLE III.** The computational efficiency of dynamic yaw error prediction.

| Case # | Testing time [s] | | | |
|---|---|---|---|---|
| | Linear | SVM | RF | Hybrid |
| 01 | 0.009 | 0.192 | 1.564 | 1.564 |
| 02 | 0.011 | 0.195 | 1.593 | 1.593 |
| 03 | 0.009 | 0.235 | 1.728 | 1.728 |
| 04 | 0.009 | 0.186 | 1.823 | 1.823 |
| 05 | 0.007 | 0.867 | 2.390 | 2.390 |
| 06 | 0.009 | 0.903 | 2.487 | 2.487 |
| Avg | 0.009 | 0.430 | 1.931 | 1.931 |

**D. Total yaw error correction**

Fig.9 shows the time series of turbine nacelle directions with and without corrections and further compares them with the real wind directions (i.e., $\theta_{WD}(t) + \theta_s$) for all six cases. In some situations, the turbine nacelle is not sensitive to tracking the high-frequency changes in wind direction associated with turbulent atmospheric flows in the field, as shown in Figs. 9(a,b,e,f). Sometimes, the turbine nacelle can orientate in time, as shown in the first 150 min in Fig. 9(c) and the first 50 min in Fig. 9(d). But sometimes, the turbine nacelle is not sensitive to the small variations in wind direction (e.g., < 6° within the 1-min interval for the Eolos turbine) due to the long wait time induced by the time tolerance auto yaw control algorithm, e.g., $t^*$ =70~120 min in Fig. 9(e). In addition, the turbine nacelle may also be stuck in a fixed position due to the yaw failure, which is reported by Pandit et al. [50] as well. No matter the situations, the predictions of dynamic yaw errors derived from ML-based algorithms can enable the turbine with instant feedback (< 2 s computational time for 200 data points as shown in Table III) to the rapid changes in the inflow structures in all situations (see corrected results in Fig. 9). The implementation of yaw error correction via introducing yaw offset to the turbine controller requires no wait time.

As listed in Table IV, all the correction methods can reduce the turbine yaw misalignment between the real wind direction and the turbine nacelle direction to some extent. Following the same trends observed for the evaluations of dynamic error prediction, all three ML-based algorithms yield higher $CF_{ye}$ than the two persistence models, PM and PM10. The performance for the total yaw error correction also ranks in order of $hybrid > RF > Linear \cong SVM$. The averaged $CF_{ye}$ for all the six cases is 71% by using the hybrid model, and such value reaches up to 85% in case#03. If the correction follows enforcement of the same wait time restriction used currently on the Eolos turbine (Clipper C96 in the present study), the average $CF_{ye}$ would decrease to 57%, with the highest value of 78% observed in case#03. Such a finding indicates that the removal of wait time may contribute to about 14% of total yaw error reduction on average.

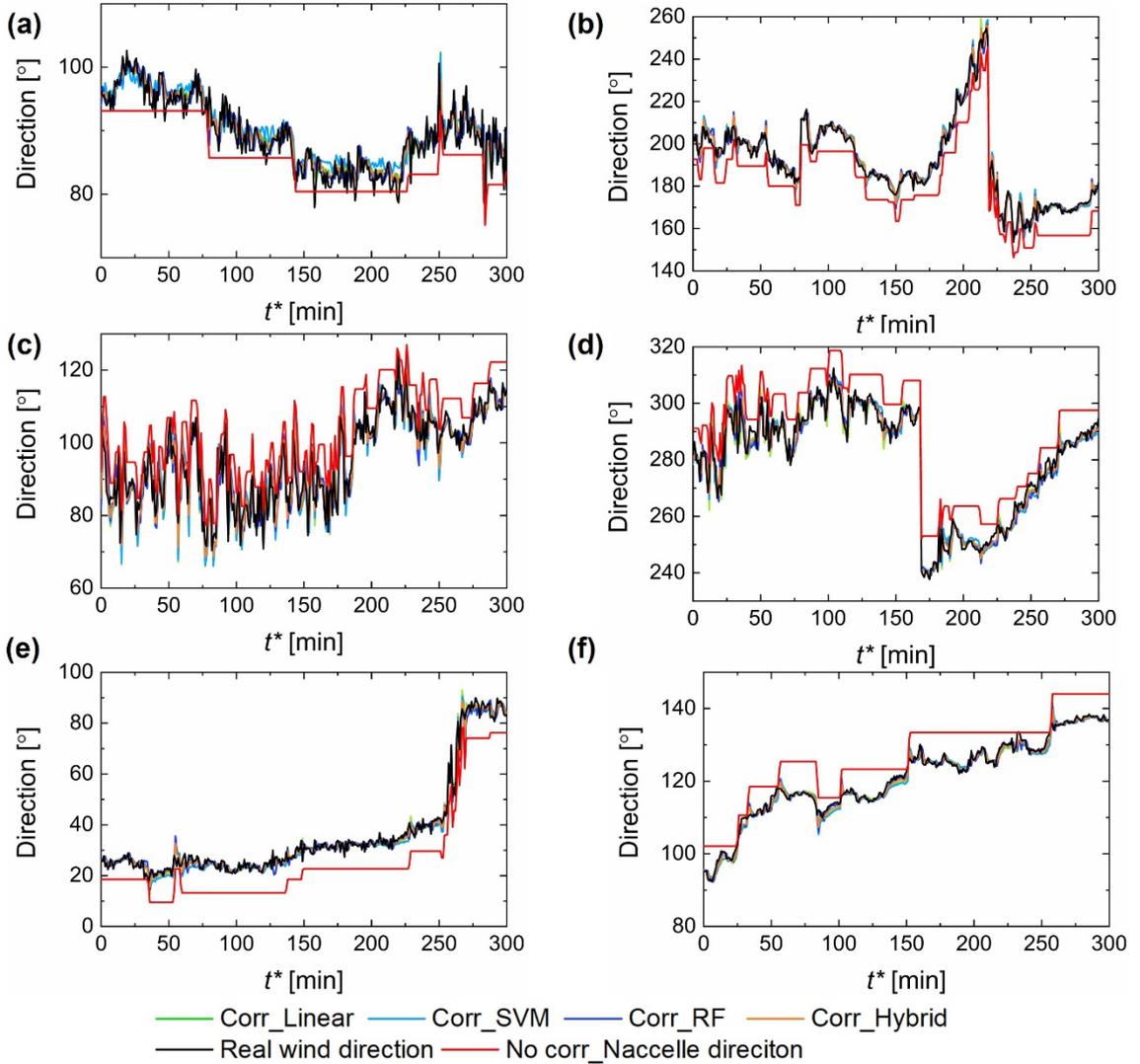

**FIG. 9.** Sample time sequences of real wind direction (i.e., $\theta_{WD}(t) + \theta_s$), measured nacelle direction ($\theta_{ND}(t)$), corrected nacelle directions using ML-based models (i.e., $\theta_{ND}(t) + \widehat{\theta_d}(t) + \widehat{\theta_s}$) in all six cases. (a) – (f) correspond to cases#01-06, respectively.

**TABLE IV**. Performance of total error correction using metric $CF_{ye}$ [%].

| Case# | Linear | SVM | RF | Hybrid | PM | PM10 |
|---|---|---|---|---|---|---|
| 01 | 54.3 | 51.0 | 54.7 | 65.8 | 45.1 | 45.4 |
| 02 | 71.3 | 70.8 | 73.0 | 79.1 | 66.8 | 61.2 |
| 03 | 78.4 | 78.3 | 81.0 | 84.7 | 74.1 | 76.3 |
| 04 | 58.9 | 59.6 | 62.6 | 71.0 | 49.9 | 53.6 |
| 05 | 58.0 | 59.0 | 61.9 | 70.5 | 49.0 | 53.1 |
| 06 | 35.7 | 37.6 | 42.0 | 54.9 | 20.9 | 33.3 |
| Avg | 59.5 | 59.4 | 62.5 | 71.0 | 51.0 | 53.8 |

### E. Transferability

Transferability describes the capability of the model derived from the benchmarking turbine to be implemented to other turbines in a wind farm. For the same type/model wind turbines, the static yaw error correction method is expected to have good transferability because the power-yaw relationship remains the same due to the identical designed aeroelastic properties and control systems. Meanwhile, the transferability of the dynamic yaw error correction method is examined for three ML-based algorithms and the hybrid model, as listed in Table IV. The two cases not included in the model training (i.e., cases#05-06 conducted with the Eolos turbine with different intended static yaw offsets) are used to represent the operational data of other turbines. These two cases exhibit effective corrections in terms of total yaw error, i.e., case#05 (hybrid:$CF_{ye}$ = 71%) and case#06 (hybrid: $CF_{ye}$ = 55%), and comparable prediction accuracy for both static and dynamic yaw errors compared with cases#01-04, implying a good transferability from the benchmarking turbine to other turbines.

### VI. SUMMARY & DISCUSSION

In the present study, we propose a framework to detect and correct the static and dynamic yaw errors for utility-scale wind turbines using the turbine SCADA records. This framework includes a data collection from the benchmarking turbine, a data mining model for static yaw error detection, and machine learning-based (ML-based) models for dynamic yaw error detection. Specifically, we first introduce offset angles to the turbine yaw controller to simulate the static yaw error conditions for the benchmarking turbine in a wind farm. At least four cases with yaw offsets ranging from -10° to 10° are suggested to ensure adequate datasets for developing the models. Such a procedure provides the ground truths of the static yaw errors without additional instruments for the wind direction measurements, such as nacelle or ground LiDARs. The collected datasets from the turbine SCADA system are then used to develop data-driven models for static and dynamic yaw error detection. The static yaw error detection mainly relies on the relationship between yaw error and corresponding power deviation associated with the yaw error-induced flow distortion. We use the arithmetic mean value of the estimates via the curving fittings on the scatters of turbine power output against yaw error in a bench of wind speed intervals to represent the final estimation of the static yaw error. In addition, we implement three ML-based algorithms, i.e., linear regression, SVM regression, and random forest (RF) regression, as well as the averaged prediction from the three tested methods (referred to as hybrid model), for the ultra-short-term (1-min ahead) forecasts of time-varying dynamic yaw errors. The determined total yaw error (i.e., the sum of static error and dynamic error) can be set into the turbine yaw controller as a correction factor to mitigate the yaw misalignment effects on turbine operation. After the data-driven models are built, only historical datasets are needed for the yaw error corrections for other turbines in the wind farm.

The proposed detection framework has been tested for a 2.5 MW wind turbine at the Eolos station from the University of Minnesota. We collect four datasets with intended static yaw offsets of 5°, 10° (positive sign: clockwise from bird's view), -8°, and -10° (negative sign: counterclockwise) to derive the power-yaw relationship for the static yaw error detection and to train the ML-based models with the first 80% of data for the dynamic yaw error prediction. The accuracy of the proposed data mining method for static yaw error detection is higher than 94% for all four cases. As to dynamic yaw error prediction, the suggested input features for training the ML-based models include turbine power, wind direction, wind speed, and five past records of dynamic yaw errors based on the grey relational analysis. For the tests using the rest 20% of data in the four cases, RF regression has the best performance, followed by SVM and linear regressions. The hybrid model can further yield a reduction in mean absolute error (MAE) in the prediction from 20% to 29%. The average total reduction in yaw misalignment for the four cases is about 75%, with a best-case of 85%. In addition, our framework has been tested using another two datasets with intended yaw offsets of 8° and -6° collected at Eolos station for the transferability analysis. These two cases show comparable accuracy in static and dynamic yaw error forecasts to the four modeling cases, demonstrating a good transferability of the proposed approach. Note that for the cases in the present study, total yaw error correction is found to be mainly from correcting the experimentally imposed static yaw error. Future work would consider the typical static yaw error in commercial turbines.

The proposed framework offers a low-cost way for yaw error detection, compared with other methods with the involvement of additional instruments for the wind direction measurements, such as nacelle or ground LiDARs. This framework only requires a simple data collection from one benchmarking turbine in a wind farm by setting intended static yaw offsets to the turbine yaw controller. Up to now, such offsets can be set to the turbine yaw controller remotely for most of the utility-scale wind turbines in the market. With the determined models, the proposed framework enables the dynamic and static yaw error detection for other turbines in the same wind farm with their historical SCADA records. Therefore, this framework is suitable for regular seasonal or annual inspections for static yaw errors in large-scale wind farms in onshore, coastal, and offshore sites. It should be emphasized that the proposed approach is more competitive and promising for offshore wind farms with significantly low accessibility for additional instrument installation due to its low cost and easy implementation. Such static yaw error detection can also be used as an initial bias inspection for wind farms that plan to improve current turbine yaw settings to minimize the turbine wake effects for global optimization of the entire farm energy production [51,52]. In addition, as for dynamic yaw error, the computational time is negligible (i.e., < 2 s for 200 data points), which allows a real-time correction for the detected errors via introducing corresponding offsets to the yaw controller. Such modification will reduce the values of yaw errors recorded in the turbine SCADA system, but it will not affect how the past observations to be used in the future prediction. But it should be cautioned that the removal of wait time for small yaw offsets (i.e., < 6° for the Eolos turbine) used in the present study may increase the yaw duty and yaw motor fatigue, which is a subject for future investigation. Moreover, as to the wind farms with more than one type of wind turbine, each type requires one benchmarking turbine for the modeling to cover the differences of the turbine-dependent features, such as blade elasticity characteristics and turbine control system properties. In future work, we will implement the proposed framework to more large-scale wind farms with online tests to further validate its generalizability.

## SUPPLEMENTARY MATERIAL

Supplementary material includes Appendix A: hyperparameter optimization of machine learning-based models, Appendix B: feature selection for the dynamic yaw error detection, and Appendix C: evidence of intended yaw offset presenting actual static yaw misalignment.

## ACKNOWLEDGMENTS

This work was supported by the National Science Foundation CAREER award (NSF-CBET-1454259), Xcel Energy through the Renewable Development Fund (grant RD4-13), as well as IonE of University of Minnesota. The authors would thank the engineers from St. Anthony Falls Laboratory, including Christopher Milliren, Chris Feist, and Matthew Lueker, for the fruitful discussion regarding the Eolos database.

## AUTHOR DECLARATIONS

**Conflict of interest:** The authors have no conflicts to disclose.

## DATA AVAILABILITY

The data that support the findings of this study are available from the corresponding author upon reasonable request.